\documentclass[useAMS,usenatbib]{mn2e}
\usepackage{graphicx,color,epsfig}

\newcommand{\be}{\begin{equation}}
\newcommand{\ee}{\end{equation}}

\newcommand{\apj}{ApJ}

\newcommand{\mnras}{MNRAS}
\newcommand{\aap}{A\&A}
\newcommand{\araa}{ARA\&A}

\newcommand{\aj}{AJ}

\def\ltsima{$\; \buildrel < \over \sim \;$}
\def\simlt{\lower.5ex\hbox{\ltsima}}
\def\gtsima{$\; \buildrel > \over \sim \;$}
\def\simgt{\lower.5ex\hbox{\gtsima}}

\def\msun{{\,{\rm M}_\odot}}

\def\del#1{{}}

\title[Dynamic Monte Carlo]{Dynamic Monte Carlo radiation transfer in SPH. \\Radiation
  pressure force implementation.}

\author[S. Nayakshin, S.-H. Cha, A. Hobbs]{Sergei
  Nayakshin\thanks{E-mail:Sergei.Nayakshin@astro.le.ac.uk}, Seung-Hoon Cha,
  and Alexander Hobbs\\ Department of Physics \& Astronomy, University
  of Leicester, Leicester, LE1 7RH, UK}

\begin{document}

\date{Received}

\pagerange{\pageref{firstpage}--\pageref{lastpage}} \pubyear{2008}

\maketitle

\label{firstpage}

\begin{abstract}
We present a new framework for radiation hydrodynamics simulations. Gas
dynamics is modelled by the Smoothed Particle Hydrodynamics (SPH), whereas
radiation transfer is simulated via a time-dependent Monte-Carlo approach that
traces photon packets. As a first step in the development of the method, in
this paper we consider the momentum transfer between radiation field and gas,
which is important for systems where radiation pressure is high. There is no
fundamental limitations on the number of radiation sources, geometry or the
optical depth of the problems that can be studied with the method. However, as
expected for any Monte-Carlo transfer scheme, stochastic noise presents a
serious limitation. We present a number of tests that show that the errors of
the method can be estimated accurately by considering Poisson noise
fluctuations in the number of photon packets that SPH particles interact with
per dynamical time. It is found that for a reasonable accuracy the momentum
carried by photon packets must be much smaller than a typical momentum of SPH
particles. We discuss numerical limitations of the code, and future steps that
can be taken to improve performance and applicability of the method.
\end{abstract}

\begin{keywords}
{Physical Data and Processes: radiative transfer -- Physical Data and
  Processes: hydrodynamics}
\end{keywords}

\section{Introduction}\label{intro}

Astronomers do not have a luxury of testing their ideas of, e.g., galaxy and
star formation in a purpose designed laboratory. Instead, testing grounds are
provided by observations and numerical simulations. The latter in effect
represent experiments with a given set of physical laws included. Clearly, the
more physics is included in the simulations the more realistic the latter should
be. Presently, two basic physical processes -- gravity and hydrodynamics --
are modelled well by a variety of methods and codes. For example, one reliable
and widely used numerical method of modelling gas dynamics is Smoothed
Particle Hydrodynamics (SPH) \citep{Gingold77,Lucy77}. It has been used in the
various fields of astrophysics, and is especially powerful when resolving
high density regions is key. The method is grid-less and fully Lagrangian in
nature, facilitating modelling of arbitrary geometry systems. For reviews of
the SPH method see, for example, \cite{Benz90,Monaghan92,Fulk94,Price04}.

Radiation transfer and interaction with matter is another basic process
operating in astrophysical systems. Starting from the pioneering ideas of
\cite{Lucy77}, great efforts have been expended to model radiation--matter
interactions in SPH.  When gaseous systems modelled are optically thick,
photons scatter or get absorbed and re-emitted multiple times before escaping
the system \citep{RybickiLightman}. This situation is well approximated by the
diffusion approximation. Several authors have already incorporated this
radiation transfer scheme in SPH
\citep{Lucy77,Brookshaw85,Whitehouse04,WhitehouseEtal05,Whitehouse06,ViauEtal06}.
The diffusion approximation is usually complemented by a flux limiter method
to model optically thin regions where photons stream freely rather than
diffuse \citep[e.g.,][]{Whitehouse04}. Most recently \cite{PetkovaSpringel08}
implemented the diffusion radiation scheme in the cosmological code Gadget
\citep{Springel05} supplementing it by the variable Eddington tensor as a
closure relation.

Ray tracing methods is a principally different approach to radiative transfer,
where the radiative transfer equation is solved along chosen directions
(rays). Most applications only consider rays that start at discrete sources of
radiation field, such as bright stars, etc., essentially neglecting the
diffuse radiation field.  As far as the density estimation is concerned, some
authors used density defined at the locations of the nearest neighbours along
the rays \citep{Kessel-Deynet00, DaleEtal07,GritschniderEtAll08}, whereas
others approximated the density field using the nodes of the tree structure
one way or another \citep{Oxley03,Stamatellos05,Susa06}. The most recent
developments use the SPH density field directly, e.g., c.f. codes SPHRAY
\citep{AltayEtal08} and TRAPHIC \citep{Pawlik08}. \cite{RI06} developed a
method to transport radiation on adaptive random lattices, and showed that the
algorith is very efficient for cosmological re-ionisation problems.

Monte-Carlo methods are similar in spirit and yet substantially different from
the ray tracing codes. In the former, the idea is to discretise the radiation
field into ``packets'', choose directions and emission time of these packets
stochastically to obey proper physical constraints, and then propagate the
packets through matter in accord with radiation transfer equations. The main
problem for the method is stochastic noise caused by photon packet statistics
\citep[for an early SPH application see][]{Lucy99}. \cite{Baes08} presents
some new interesting ideas about using a smoothing kernel in a Monte-Carlo
simulation.


Most of these applications were tailored to photo-ionisation problems in the
field of star formation or cosmology, where the density field can be
considered static as rays propagate through it \citep[the ``static diffusion
limit''; see ][]{KrumholzEtal07}, and where radiation pressure effects can be
omitted. There are also a number of radiation transfer codes working with a
grid-based hydrodynamics codes rather SPH \citep[e.g.,][]{IlievEtal06}. We do
not discuss these methods here.

Here we present and test a new photon packet based radiation transfer scheme
combined with an SPH code. The method uses the SPH density field directly,
thus preserving the ``native'' SPH resolution. The previous efforts did not
consider the radiation pressure effects, studying instead radiative
heating/cooling and photo-ionisation. Here we study the radiation pressure
effects instead. We assume that gas equation of state is known (e.g.,
isothermal or polytrophic) and consider only the radiation pressure
forces. This allows us to thoroughly test precision of our approach and
numerical noise effects. There is no fundamental difficulty in including the
heating/cooling and photoionisation processes in our scheme, and we shall
extend our method in that direction in the near future.  Another defining
characteristic of our new method is that the photon field is evolved in the
same time-dependent way, although on shorter time steps, as gas dynamics, and
therefore the method is intrinsically time-dependent.

\del{In $\S$\ref{sec:approach} the method is described in detail. Tests to
check the exact balance between the radiation and gravitational
accelerations are presented in 
$\S$\ref{sec:single} and $\S$\ref{sec:static}. The accretion tests
will be given in $\S$\ref{sec:accretion} and $\S$\ref{sec:moving}.
In $\S$\ref{sec:conclusion}, summary and further developments will be
discussed.}

\section{Description of the method}\label{sec:approach}

\subsection{Radiation Transfer Method}\label{sec:SMC}

Radiative transfer equation along a ray \citep{RybickiLightman} is
\begin{equation}
\frac{1}{c}\,\frac{\partial I}{\partial t} = -  \kappa \rho I + \varepsilon\;,
\label{perenos}
\end{equation}
where $I$ is the specific radiation intensity, $\kappa$ is the opacity
coefficient, $\rho$ is gas density and $\varepsilon$ is emissivity of the
gas. In general $I$ and $\varepsilon$ are functions of direction, radiation
frequency, position and time. The first term on the right hand side represents
removal of radiation from the beam by absorption and scattering, whereas the
last term describes local emission of radiation into the beam's direction.

In Monte Carlo methods, the radiation field is sampled via photon packets.  In
the simplest reincarnation of the method, both terms on the right hand side of
equation \ref{perenos} are treated stochastically. The photon mean free path,
$\lambda$, is calculated as $\lambda = 1/(\kappa \rho)$. A random number,
$\xi$, uniformly distributed over $[0,1]$, is generated. A photon is allowed
to travel distance $\Delta l = - \lambda \ln\xi$, at which point it interacts
with gas by passing its energy and momentum to gas {\em completely}. The
photon is then re-emitted according to physics of a chosen set of radiation
processes, and followed again until it escapes the system.

Our radiation transfer scheme is slightly modified from this. Firstly, instead
of using discontinuous photon jumps, we explicitly track packet's trajectory
in space as
\begin{equation}
{\bf r}(t) = {\bf r_0} + {\bf v_\gamma} t
\end{equation}
where $\bf{r}(t)$ and ${\bf r_0}$ are the current and initial photon
locations, and $|{\bf v_\gamma}|=v_\gamma =$~const is the photon propagation
speed (not necessarily the speed of light; see \S \ref{sec:prompt}).
Secondly, photon momentum, $p_\gamma$, and photon energy $E_{\gamma} =
cp_\gamma$ are reduced {\em continuously} due to absorption or scattering as
\begin{equation}
\frac{1}{v_\gamma}\frac{dp_\gamma}{dt} = - \frac{p_\gamma}{\lambda}\;.
\label{absorption_only}
\end{equation}
Thus, the first term on the right hand side of equation \ref{perenos} is
treated continuously rather than stochastically.  This reduces the statistical
noise significantly. Photons are discarded when their momentum drops below a
small ($\sim 10^{-4}$) fraction of their ``birth momentum'' $p_{\gamma 0}$.

However, the re-emission term, i.e., the last one in equation \ref{perenos} is
modelled similarly to classical Monte-Carlo methods in that it is stochastic
in nature. This is unavoidable in Monte Carlo methods due to the need to
employ a finite number of photon packets \citep[e.g.,][]{Lucy99}.  In
practice, at every photon's time step $\Delta t_\gamma$, the photon momentum
absorbed, $\Delta p_\gamma$, is calculated according to equation
\ref{absorption_only}. As stated in the Introduction, we concentrate in this
paper on the radiation pressure effects, and assume that radiation absorbed
from the beam is completely re-emitted on the spot in a new random
direction. This is the case for pure scattering of radiation or for local
radiative equilibrium between radiation and gas. In these cases the amount of
momentum (energy) to be re-emitted in the interaction considered is exactly
$\Delta p_\gamma$.

Probability of re-emitting a new photon with momentum (energy) $p_{\gamma 0}$
($c p_{\gamma 0}$)is defined as $w_\epsilon = \Delta p_\gamma/p_{\gamma 0}$. A
random number, $\xi$, uniform on $[0,1]$, is generated. If $\xi < w_\epsilon$,
a new photon with momentum $p_{\gamma 0}$ and a random direction of
propagation is created. This conserves photon field's energy and momentum in a
time-average sense.

This scheme can be adapted to allow for non-equilibrium situations when
radiative cooling is not equal radiative heating. Multi-frequency radiation
transfer is also straight forward if tedious to implement. We have done some
of these developments already and will report it in a future paper.

\begin{figure*}
\begin{minipage}[h]{.48\textwidth}
\centerline{\psfig{file=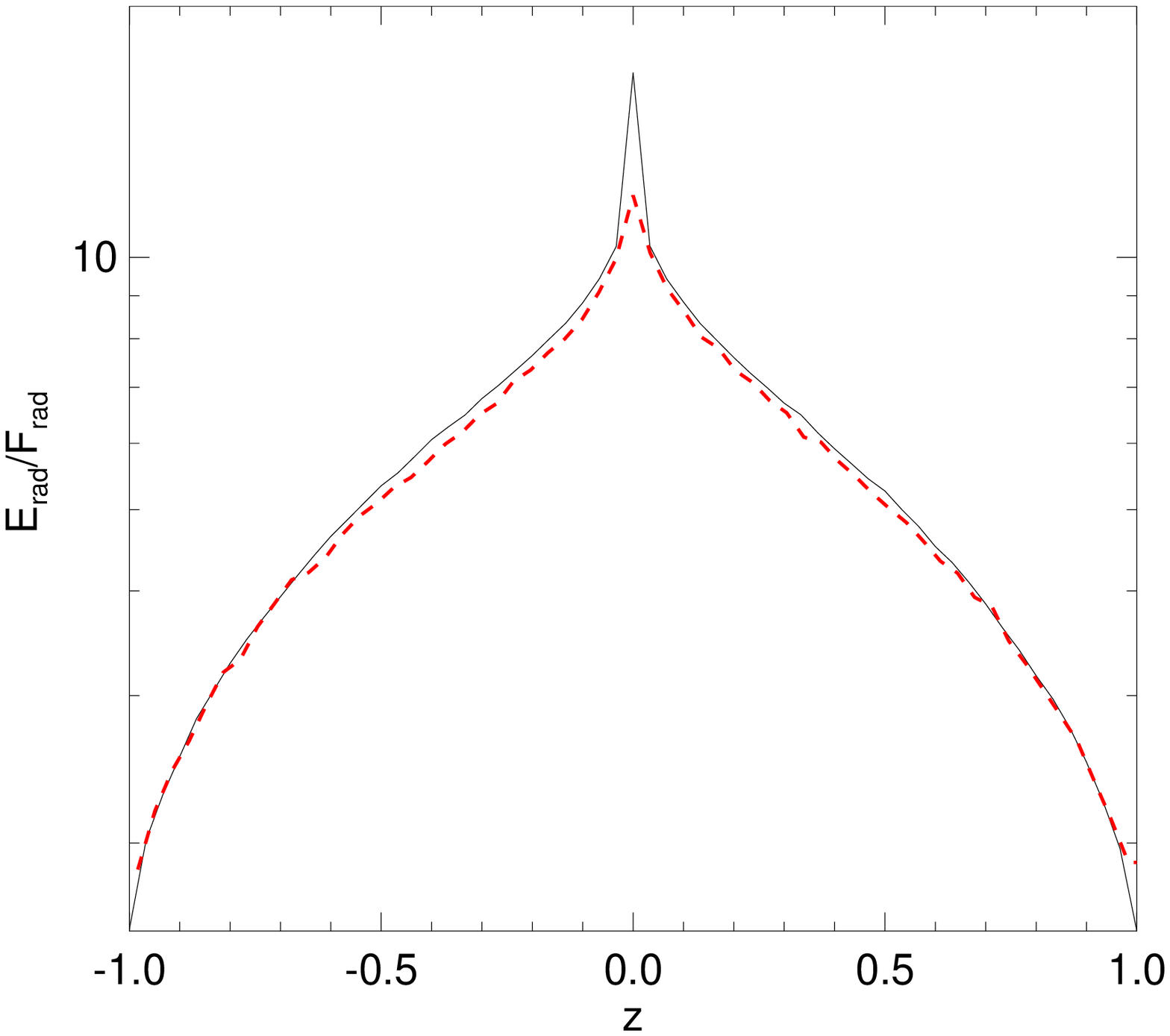,width=0.99\textwidth,angle=0}}
\end{minipage}
\begin{minipage}[h]{.48\textwidth}
\centerline{\psfig{file=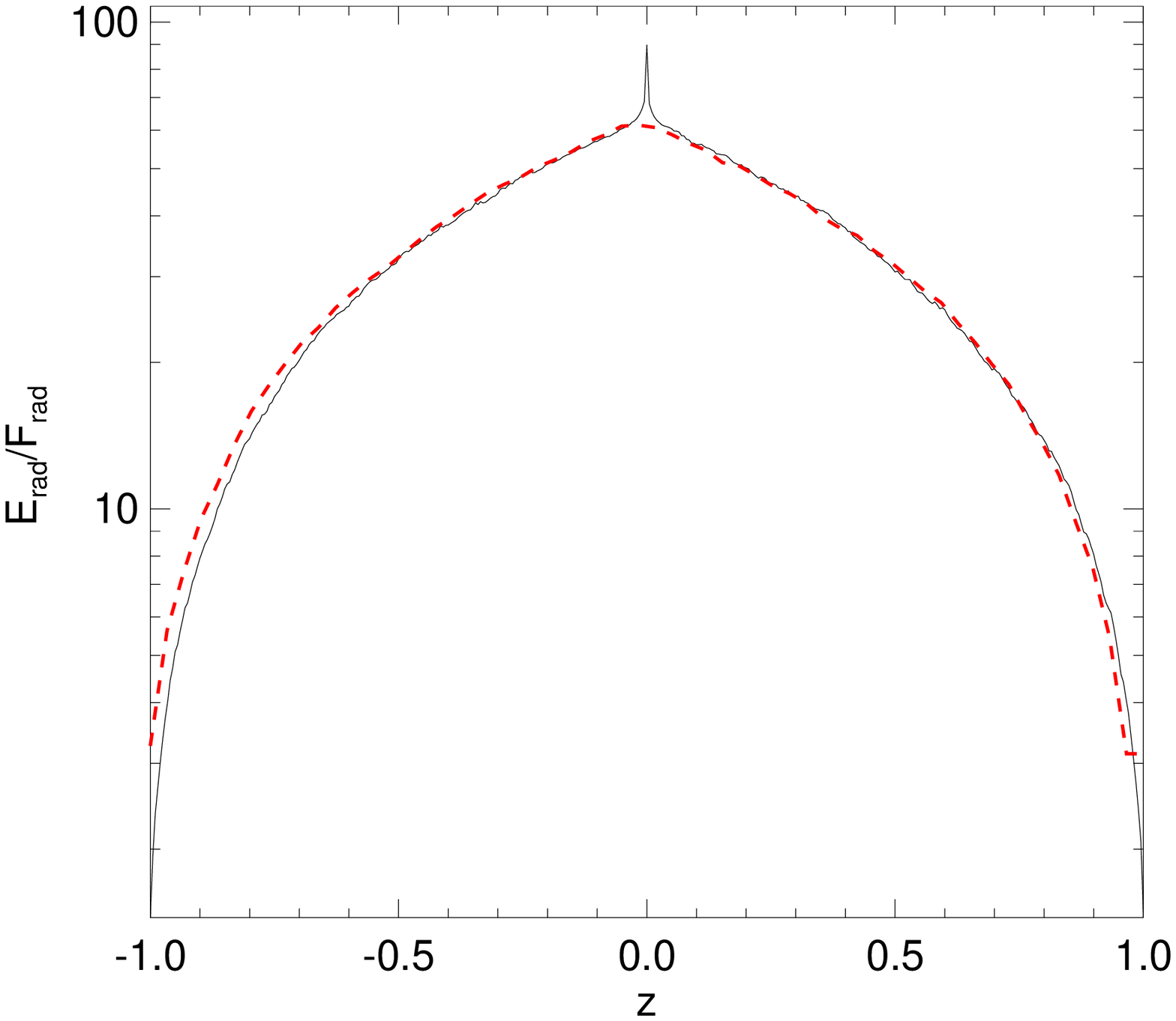,width=0.99\textwidth,angle=0}}
\end{minipage}
\caption{Comparison of radiation energy density inside a slab of matter
  computed via our radiation transfer formalism (red dashed curves) and by the
  traditional Monte Carlo technique (solid). The difference is minor, and is
  entirely due to a finite time-resolution of the red curve, as explained in
  the text.}
\label{fig:slab_rt}
\end{figure*}

\subsection{A static test of radiation transfer}\label{sec:slab_transfer}

The radiation transfer approach described in \S \ref{sec:SMC} embodies the
classical radiation transfer equations and must be correct on that
basis. Nevertheless, we tested our approach against the discrete Monte-Carlo
method on a simple test problem where the gas density and opacity are fixed
and independent of the radiation field. This setup does not require using gas
dynamics at all, and hence no SPH particles were used to model the density
field. In the test, photons are emitted at a constant rate at the midplane of
a plane-parallel slab $z = [-1, 1]$. The main parameter of the problem is the
total optical depth of the gas between $z=0$ and $z=1$, $\tau_t$. We ran the
radiation transfer part of the code until a steady state photon packet
distribution was achieved. 

Figure \ref{fig:slab_rt} shows the ratio of the local radiation energy density
to the radiation flux out of the slab as a function of $z$-coordinate inside
the slab. For convenience, the speed of light is set to unity. Two tests are
shown, one for $\tau_t=2$ (left panel) and the other for $\tau_t = 20$ (right
panel). Black solid curves show the same quantity computed with the standard
Monte Carlo technique, whereas the red dashed curves are computed with the
modified method described above. For the latter time-dependent calculation, we
averaged over a number of snapshots. The disagreement is quite small
everywhere except for the region near $z=0$. The difference in that region is
simply due to the finite time resolution of the red curves. Namely, these
curves are obtained by averaging snapshots of the simulation data. The
time step between the snapshots is long enough for the photons introduced at
times between the snapshots to diffuse away from their starting $z=0$
location, broadening the peak. The Monte Carlo curve, on the other hand, is
obtained by following individual photons' progress out of the slab, and
therefore the information about the initial location of the photon is retained
and is evident in the peak at $z=0$. These tests demonstrate that the
radiation transfer is modelled properly with our approach at least in the
simple situations considered here.

\subsection{Interactions between photon packets and SPH
  particles}\label{sec:interactions} 

We now discuss coupling the radiation transfer scheme to SPH.  A photon mean
free path is given by
\begin{equation}
\lambda = \frac{1}{\kappa \rho}\;,
\label{lambda}
\end{equation}
where $\kappa$ is the gas opacity coefficient and $\rho$ is gas mass density
{\em at the photon's current location.} We require an accurate method of local
gas density estimation, and therefore we use directly the density field
representation of the SPH method.  In the SPH, gas density at a position ${\bf
  r}$ is given by the superposition of smoothed contributions from individual
SPH particles:
\begin{equation}
\rho({\bf r}) = \sum_i m_i W\left( |{\bf r - r_i}|, h_i\right) = \sum_i \rho_i({\bf r})\;,
\label{eqn:density}
\end{equation}
where $W$ is the SPH kernel, ${\bf r_i}$ is the location, $h_i$ is the
smoothing radius of particle $i$, and $\rho_i \equiv m_i W_i$ is the
contribution of particle $i$ to the density at ${\bf r}$. The summation goes
over all the ``neighbours'' -- SPH particles that contribute to $\rho$ at
${\bf r}$, i.e., have non-zero values of $W$ at this location. Note that in
this approach there is no expectation of the number of gas neighbours of a
photon, $N_{\rm gn}$, to be constant or limited in any other way. In
particular, having $N_{\rm gn} = 0$ is perfectly feasible and simply means
that the photon propagates through an ``empty'' patch of space.

Having determined $\lambda$ at the photon's location, we find the decrement in
photon's momentum, $\Delta {\bf p}_\gamma$, at that location according to
equation \ref{absorption_only}. This decrement is then passed to the photon's
SPH neighbours to enforce conservation of momentum. If $N_{\rm gn} > 1$, there
is a question of what fraction of $\Delta {\bf p}_\gamma$ should be passed to
a particular neighbour $i$. We assume that the neighbours contribute to the
interaction with a photon directly proportionally to their density $\rho_i$ at
the photon's location (see equation \ref{eqn:density}). Hence the momentum
passed to neighbour $i$ is
\begin{equation}
\Delta {\bf p}_{\gamma i}= \frac{ \rho_i({\bf r})}{\rho({\bf r})} \Delta
{\bf p}_\gamma\;.
\label{deltap}
\end{equation}
As we limit the photon time step to ensure that $\Delta \tau = v_\gamma \Delta
t_\gamma/\lambda \ll 1$ (see equation \ref{dtgamma}), this approach yields
physically correct result.

The momentum transferred to the SPH particle $i$ from interactions with
different photon packets $\gamma$ is additive, and is used to define the
radiation pressure force on that particle. We sum all the interactions that
the SPH particle experienced during its time step $\Delta t$, and then define
the radiation pressure force on particle $i$ by
\begin{equation}
{\bf f_{{\rm rad}, i}}
= \frac{\sum_\gamma \Delta {\bf p}_{\gamma i}}{\Delta t} \;.
\label{frad}
\end{equation}

Due to limitations on photon propagation time steps $\Delta t_\gamma$ (see
below), an SPH particle may interact several times with a particular photon
$\gamma$. There is however ho paradox here, as this simply means that the
radiation transfer equation is integrated in multiple points along the
photon's trajectory within each SPH particle, increasing the precision of the
method over traditional Monte-Carlo approaches.  This formulation also allows
us to enforce an exact pair-wise conservation of momentum in the interactions
of matter and radiation. The energy transfer can be calculated in exactly same
manner except $\Delta {\bf p}_\gamma$ is replaced by $\Delta E_{\gamma}$.

\subsection{Photon packet creation and propagation}\label{sec:propagation}

External radiation sources such as stars or accreting compact objects emit
photon packets with a given momentum $p_{\gamma 0}$.  The rate of packet
emission is given by
\begin{equation}
\dot N_{\gamma} = \frac{L}{c p_{\gamma 0}}\;,
\label{ngamma}
\end{equation}
where $L$ is the luminosity of the source and $c$ is the speed of light. The
radiation field can be chosen to be isotropic or beamed/restricted to a range
of directions.

There are several issues to consider when deciding how far the photon can
travel in a single flight $\Delta l = v_\gamma \Delta t_\gamma$. First of all,
this distance should be much smaller than the typical SPH smoothing length,
$h$, in the region, or else the photon will ``skip'' interactions with some
SPH particles altogether. This constraint is important in both optically thin
and thick regimes. In the optically thick case, an additional constraint need
to be placed to insure that photons do not propagate in one step by more than
a fraction of their mean free path, $\lambda$. In the opposite case photons
would ``diffuse'' though optically thick regions in an unphysical way, i.e.,
too quickly. These constrains are combined by requiring
\begin{equation}
\Delta t_\gamma = \delta_t \hbox{min}\left[\frac{h_\gamma}{v_\gamma}\,,
 \frac{\lambda}{v_\gamma} \right]\;,
\label{dtgamma}
\end{equation}
where $\delta_t \ll 1$ is a small dimensionless number. In practice we use
$\delta_t = 0.03 - 0.3$.

As discussed below in \S \ref{sec:prompt}, in the ``prompt escape'' regime, it is
possible to reduce the photon propagation speed below the speed of light,
which then allows us to integrate photon trajectory on longer time steps
without compromising the physics of the problem. 

\subsection{Regimes of radiation transfer}\label{sec:prompt} 

Following \cite{KrumholzEtal07}, radiation hydrodynamics of a problem can be
divided into three different regimes. Let $u$ be a characteristic gas
velocity, such as the sound speed or the bulk gas velocity, whichever is
greater. Define $\beta = u/c$ and optical depth of the system $\tau =
l/\lambda$, where $l$ is the geometric size of the system. The three limiting
regimes \citep{KrumholzEtal07} are
\begin{equation}
\tau \ll 1 \qquad \hbox{the free streaming limit,}
\label{free}
\end{equation}
\begin{equation}
\tau \gg 1, \quad \beta\tau \ll 1 \qquad  \hbox{the static diffusion limit,}
\label{static}
\end{equation}
\begin{equation}
\tau \gg 1, \quad \beta\tau \gg 1 \qquad  \hbox{the dynamic diffusion limit.}
\label{dynamic}
\end{equation}

In the first regime a typical photon leaves the system in a single flight. In
the second case the photon scatters or get absorbed and re-emitted
approximately $\tau^2 \gg 1$ times before leaving the system. For what follows
it is important that in the first two regimes photons escape from the system
on timescale, $t_{\rm esc}$, much shorter than the matter distribution can
alter significantly, i.e.,
\begin{equation}
t_{\rm esc} = \frac{R}{c}(1 + \tau) \ll \frac{R}{u}\;.
\label{tesc}
\end{equation}
Here $R$ is the geometric size of the system.  In contrast, in the dynamic
diffusion limit the photon diffusion time is larger than the crossing time of
the system, $R/u$.

Equation \ref{tesc} shows that the exact value of the speed of light is
irrelevant in the free streaming and the static diffusion limits; it is so
high it can be considered infinite. To the contrary, in the dynamic diffusion
limit the exact value of speed of light is important as it defines the time
scale on which radiation from the system leaks out, and that time scale is
long.

Therefore, we combine the free streaming limit and the static diffusion limit
of \cite{KrumholzEtal07} into the ``prompt escape'' regime given by the
equation \ref{tesc}. In this limit it is numerically convenient and physically
permissible to reduce the photon propagation speed $v_\gamma$ below $c$, as
long as the system still satisfies equation \ref{tesc} with $c$ replaced by
$v_\gamma$.  We found that this speeds up calculations in which gas dynamics
is important, although the scaling is not as efficient as the factor
$c/v_\gamma$. The reason for that is that although photon time step (equation
\ref{dtgamma}) is indeed longer by the factor of $c/v_\gamma$, the number of
photon packets for a given $p_\gamma$ is correspondingly higher (see equation
\ref{ngamma} and note that $N_\gamma \sim \dot N_\gamma t_{\rm esc}$).

\subsection{Implementation in Gadget}\label{sec:gadget}

This radiation-gas momentum transfer method has been implemented in the
SPH/N-body code Gadget \citep{Springel05} that is widely used for cosmological
simulations. For the tests presented here the cosmological options of the code
are turned off. Gadget uses a Barnes-Hut tree to speed up calculation of
gravitational forces and for finding neighbours. As photon packets have no
mass associated with them, we turn off the gravity calculation for these
particles. They are also not included in building the Barnes-Hut tree.

To find SPH neighbours of a photon, we first find all SPH particles that are
inside a sphere with size $h_{\rm search}$ which is chosen to be much larger
than the mean SPH particle smoothing length. We then further select only those
SPH particles $i$ that contain the photon within their smoothing length $h_i$.

After calculating the radiation pressure force for a given SPH particle $i$,
the corresponding radiative acceleration ${\bf a_{{\rm rad}, i}} = {\bf
  f_{{\rm rad}, i}}/m_i$ is added to the hydrodynamical and gravitational
accelerations that the particle experiences. We have also added the radiation
pressure acceleration to the time step criteria for the SPH particles, as
described in \cite{Springel05}. This ensures that SPH particle time steps are
appropriately short for particles with large radiation pressure accelerations.

\subsection{Static SPH radiation transfer test}\label{sec:static_disk}

In \S \ref{sec:slab_transfer} we tested the radiation transfer methods for the
density field given by a simple analytical function (a constant). In \S
\ref{sec:interactions} we presented a way to model radiation transfer in an
arbitrary density field represented by the SPH particles. It is a logical step
forward in complexity of the tests to now repeat the slab test of \S
\ref{sec:slab_transfer} in a self-consistent SPH density field.

To accomplish this, we consider a non self-gravitating accretion disc in orbit
around a massive ($M=1$ in code units) central source. The disc is assumed to
be locally isothermal, with internal energy $u$ scaling as $u = u_0 (R_0/R)$,
where $R$ is radius, and $u_0=0.02$ is the internal energy at the inner edge
of the disc, $R_0 = 1$. This scaling yields a constant ratio of the vertical
disc scale height, $H$, to radius. The outer radius of the disc is $R_{\rm out}
= 3$. The disc was then relaxed for a large number of orbits without any
radiation field, which yields a Gaussian vertical density profile
\citep{Shakura73}. Keeping this density profile fixed, we then introduced
photon sources in the disc midplane and allowed the photons to propagate out
of the disc keeping the opacity coefficient $k$ fixed everywhere in the disc.

\begin{figure}
\centerline{\psfig{file=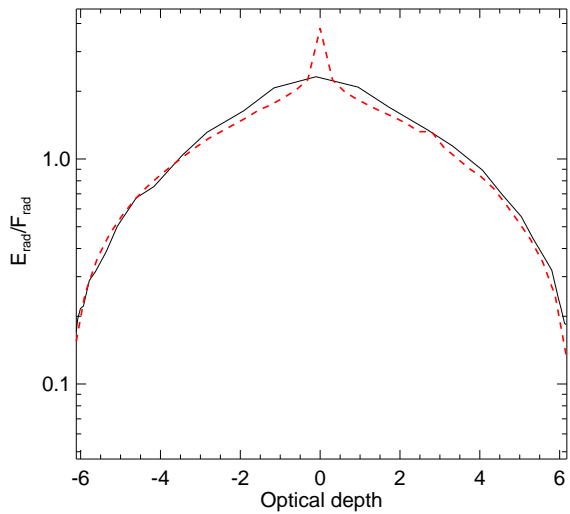,width=0.49\textwidth,angle=0}}
\caption{The ratio of the radiation energy density to the radiation flux for
  the simulation described in section \ref{sec:static_disk} (black solid
  curve) compared with a traditional Monte-Carlo calculation for a slab with
  same optical thickness (red dashed curve). While similar in spirit to tests
  shown in Figure \ref{fig:slab_rt}, the present simulation uses an actual SPH
  density from a live accretion disc simulation.}
\label{fig:slab_transfer}
\end{figure}

Figure \ref{fig:slab_transfer} shows the resulting photon energy density
distribution within the slab (solid curve) in exactly same manner as in the
fixed density tests shown before in Figure \ref{fig:slab_rt}. As before, the
radiation field has been averaged over several snapshots to reduce statistical
noise. The respective Monte-Carlo result obtained for the same density
distribution is shown in Figure \ref{fig:slab_transfer} with the red dashed
curve. There is a very good agreement between the curves, except for the peak
region in the red curve. Time sampling differences in the two simulations
explain the discrepancy in the curves; see discussion of Figure
\ref{fig:slab_rt} in \S \ref{sec:slab_transfer}. This test suggests that the
radiation transfer part of our approach is working as expected. Below we shall
move on to tests that involve gas dynamics.

\section{Absorption on the spot tests}\label{sec:onspot}

Young massive stars produce most of their radiation in the UV
wavelengths. Dust in the interstellar gas has a very large absorption opacity
for UV photons, with $k_{\rm uv}$ up to a few hundred cm$^2/$ g. If these
photons are absorbed and re-emitted in the infrared, where Roseland opacity
is orders of magnitude smaller than $k_{\rm uv}$, in many applications one can
effectively assume $k_{\rm ir} = 0$. In this case it is sufficient to consider
only the radiation pressure from the UV photons emitted by the young stars,
and neglect the re-radiated component. Finally, one can approximate the large
UV opacity by an infinitely large one, $k_{\rm uv} = \infty$.  Photon packet
propagation is then trivial: packets travel in straight lines with constant
momentum $p_{\gamma}$ until they encounter an SPH particle(s), at which point
they are absorbed and their momentum is transferred to that particle(s).

This ``on the spot'' absorption method can also be used to model {\em fast}
gas winds from massive stars or luminous black holes in the momentum driven
regime. In the latter case the cooling time in the outflow is short. Shocked
outflow gas cools very quickly, and hence its thermal pressure can be
neglected. The momentum transferred to the ambient medium provides the push to
drive the shell out. If the wind velocity is much higher than the velocity of
the expanding shell and the speed of sound in the ambient gas, one can 
also neglect the mass outflow from the source. This is justified as the mass
flux in the wind is small compared with that of the ambient gas being driven
out.

The ``on the spot'' approximation presents a convenient test ground of our
code since it is possible to derive exact analytical solutions in the simplest
cases. In particular, we consider a single radiation source embedded in an
infinite initially uniform isothermal medium. Gravity is turned off for
simplicity.

In practice we set up periodic boundary conditions for a cubic box with
dimensions, $l$, of unity on a side. These boundary conditions are appropriate
and do not affect our results since the radiation force effects are contained
to a small region within the box during the simulations. Gas internal energy
is fixed at $u=1$, hence sound speed $c_s = 1$. The total mass of the gas
inside the box is $M=1$. The unit of time is $l/c_s$. $10^6$ SPH particles is
used in these tests. The initial condition is obtained by relaxing the box
without the radiation source to a state of nearly constant gas density over
hundreds of dynamical times for the simulation box.

\subsection{A steady-state case}\label{sec:steady}

\begin{figure*}
\centerline{\psfig{file=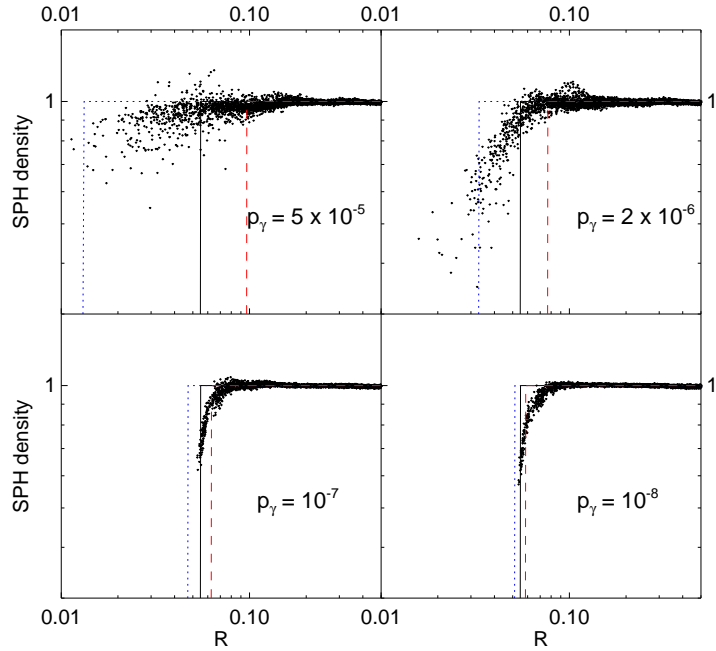,width=0.6\textwidth,angle=0}}
\caption{ Density of selected SPH particles versus radial distance
  from the source for tests described in \S \ref{sec:onspot}. The
  dimensionless luminosity of the source is $L_{\gamma} = 0.025$ for all the
  tests. The momentum of the individual photon packets varies between
  $p_\gamma = 5 \times 10^{-5}$ to $p_\gamma = 10^{-8}$, as indicated in each
  of the panels. The solid line shows theoretically expected density profile,
  whereas the dotted and the dashed lines show the estimated error.}
\label{fig:onthespot}
\end{figure*}

\begin{figure*}
\centerline{\psfig{file=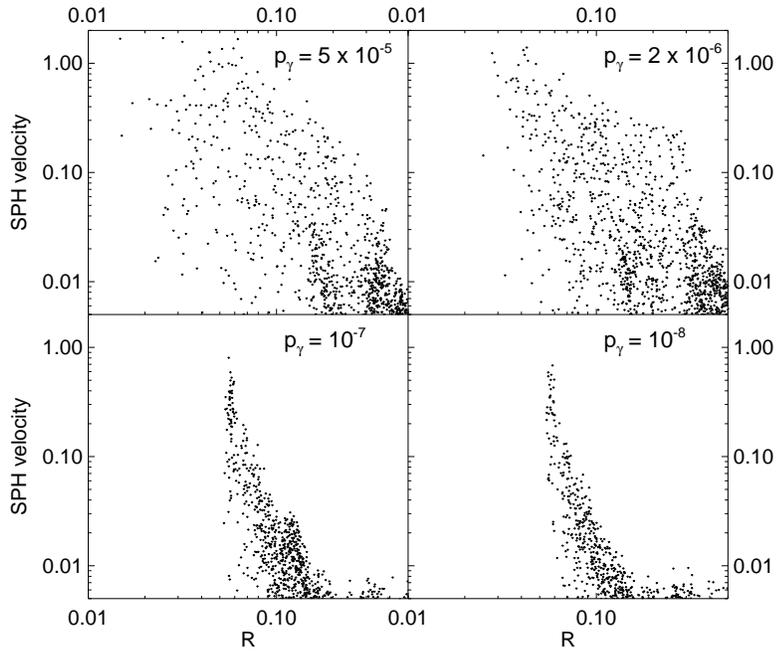,width=0.6\textwidth,angle=0}}
\caption{Absolute velocity of SPH particles for tests shown in Figure
  \ref{fig:onthespot}. Theoretically expected solution would have $v = 0$
  everywhere. As before, the results in the two upper panels, corresponding to
  the cases of a large photon momentum, are quite inaccurate and show
  significant numerical fluctuations in $v$. The lower panels show such
  fluctuations only in the narrow layer adjacent to the inner
  cavity. Fluctuations behind the layer are very subsonic and can be neglected
  for practical purposes.}
\label{fig:onthespot_v}
\end{figure*}

We first consider a case with a relatively small source luminosity $L_{\gamma}
= 0.025$ in the code units. In this case a quasi-steady state should be set up
quickly. We consider that state here. Figure \ref{fig:onthespot} shows the
density profiles obtained in four different runs at dimensionless time $t =
3.5$, when the SPH quantities approach a quasi steady state. Theoretical
expected density profile is given by $\rho = 0$ for $R \le R_{\rm cav}$ and
$\rho = $~const~$\approx 1$ for $R \ge R_{\rm cav}$. Here $R_{\rm cav}$ is the
size of the cavity opened up by the radiation. This is obtained by requiring a
force balance between the momentum flux from the source, $L_{\gamma}/c$, and
the external pressure of the gas $p_{\rm ext}$:
\begin{equation}
R_{\rm cav} = \left[\frac{L_\gamma}{4\pi c p_{\rm ext}}\right]^{1/2}\;.
\label{rcav}
\end{equation}
The expected discontinuous density profile is shown with the solid (black)
line (we have neglected a slight density increase due to evacuation of the gas
from the cavity, since $R_{\rm cav} \ll l$, the box size). Figure
\ref{fig:onthespot_v} shows instantaneous velocities of SPH particles
corresponding to the respective panels in Figure
\ref{fig:onthespot}. Theoretically expected result for velocity in the steady
state is $v=0$ everywhere, of course.

Figures \ref{fig:onthespot} and \ref{fig:onthespot_v} show that at relatively
large values of the photon packet momentum, corresponding to a lower packet
injection rate $\dot{N}_\gamma = L_\gamma/(c p_\gamma)$, the method is
inaccurate. For $p_\gamma = 5 \times 10^{-5}$, no cavity is opened at all
around the source. Moreover, SPH particle velocities have random components at
a 30\% fraction of the sound speed, even at regions far behind the expected
discontinuity at $R = R_{\rm cav}$. In contrast, tests with $p_\gamma =
10^{-7}$ and $10^{-8}$ produce nearly identical results with velocity
fluctuations below a few percent level at $R \simgt 2 R_{\rm cav}$.

Physically, if $p_\gamma > p_{\rm sph}$, interaction of an SPH particle with a
single photon packet may accelerate the particle to velocity exceeding the
sound speed, leading to a significant numerical noise. In the tests presented
here, a typical SPH particle momentum is $p_{\rm sph} = m_{\rm sph} c_s =
10^{-6}$, as $ m_{\rm sph} = 10^{-6}$. Therefore, the two runs with $p_\gamma
> p_{\rm sph}$ do very poorly, as expected.  These numerical experiments show
that the minimum accuracy requirement {\em in optically thick} limit is
\begin{equation}
p_\gamma \simlt p_{\rm sph}\;.
\label{psph}
\end{equation}

For further quantitative analysis of the errors, consider now the width
$\Delta R$ of the transition layer. This layer is defined as the region in
which SPH density goes from zero to unity. First, note that on average, a gas
parcel within volume $\sim h^3$ situated on the inner face of the transition region
receives one photon packet every
\begin{equation}
\Delta t_1 = \frac{1}{\dot{N}_\gamma}\frac{4 \pi R_{\rm cav}^2}{\pi h^2}\;
\label{deltat1}
\end{equation}
seconds. During this time, SPH particles inside the parcel experience no
radiation pressure ``pushes''. We assume that in the absence of radiation
forces, the volume is quickly accelerated by the pressure gradient force to
$v_R \approx - c_s$. There are $\sim N_{\rm nb}$ SPH particles within the
volume, where $ N_{\rm nb}$ is the typical number of SPH neighbours (usually
chosen to be around 40). Due to the spherical symmetry of the problem, the
parcel travels inwards to the radiation source a distance $\Delta R = v_R
\Delta t_1$ before it is turned back by the arrival of the next photon packet.
Finally, since the volume element is located on the inner boundary of the
cavity evacuated by the radiation pressure, it will expand when moving into
the cavity. Hence the appropriate smoothing length of the element will be
larger than that far from the cavity, and in general we expect $h \sim \Delta
R$ in that region. Using the latter estimate in equation \ref{deltat1}, we
obtain for the thickness of the transition layer,
\begin{equation}
\frac{\Delta R}{2 R_{\rm cav}} = \left[\frac{c_s}{2 R_{\rm cav} \dot{N}_\gamma }\right]^{1/3}\;.
\label{deltar}
\end{equation}
The transition layer thickness calculated in this way is indicated in Figure
\ref{fig:onthespot} with dotted ($R_{\rm cav} - \Delta R_{\rm tr}$) and dashed
($R_{\rm cav} + \Delta R_{\rm tr}$) lines. It appears to be a good estimate
for the three runs out of the four shown in the Figure. In the run with the
highest photon packet resolution, i.e., the smallest $p_\gamma$, equation
\ref{deltar} is an under-estimate. This is simply because a density
discontinuity in SPH cannot be narrower than the width of the kernel, i.e.,
the minimum smoothing length, which is about $h = 0.02$ for all the four
tests.  Inserting equations \ref{rcav} and \ref{ngamma} into equation
\ref{deltar}, and requiring ${\Delta R}$ to not exceed the characteristic
smoothing length of the problem leads to the same requirement of the packet's
momentum to be small enough compared with the typical SPH momentum (equation
\ref{psph}).

Summarising the results of these tests, we see that we can use the condition
\ref{psph} as the photon packet resolution requirement in an optically thick
limit. In addition, stochastic noise arguments appeared useful when
determining the width of the transition layer in Figure \ref{fig:onthespot}.


\subsection{Momentum-driven wind test}\label{sec:wind}

Keeping to the model of a single source radiating photons isotropically into
an infinite, constant density, isothermal medium, we performed a further test
whereby the luminosity of the source was significantly higher, $L_{\gamma} =
50$. This time the focus of our attention is the evolution of the radius of the expanding shell with time,
well before it reaches the equilibrium cavity size.

This test is closely related to the ram-pressure driven wind test
\citep[e.g.,][]{vishniac1983, garcia-segura1996, DaleandBonnell08}, but with the important
distinction that in our method the photon `gas' is massless. This however
should be a good approximation to a low density but high velocity outflow.

The other important aspect of our model is the isothermal condition,
corresponding physically to a shocked gas region that is allowed to cool very
quickly, i.e. when the cooling time in the shocked gas is far shorter than the
dynamical time. This tends to occur when the temperature is less than $10^6$
K, so that line cooling from ions dominates \citep{stellarwindsbook}. The
picture then is that of a momentum-conserving `snowplow' phase of an expanding
gaseous bubble as it sweeps up the ambient gas into an increasingly massive
shell. This model has three zones: (i) an evacuated cavity through which
free-streaming photons are being emitted from the central star (ii) a shell of
shocked ISM gas at temperature $T_{0}$ bounded by an outer shock front and
(iii) the undisturbed uniform ISM at temperature $T_{0}$.

The initial conditions for this test and numerical setup are same as in \S
\ref{sec:steady}. The choice of the higher luminosity is motivated by a
requirement that the Mach number be high so that the pressure from the ambient
gas did not have a significant effect on the evolution of the shell during the
run. To aid this, the gas internal energy was set to a lower value of $u =
0.1$.


The analytical solution to a momentum-conserving bubble can be derived by
considering the thin-shell approximation, whereby the swept-up gas is assumed
to be concentrated in an infinitely thin shell that is being driven by the
impinging wind. In reality the shell gets thicker the more gas it sweeps up,
but at all times the thickness of the shell, $D$, is much less than the radius
\citep{clarkeandcarswell03}:
\begin{equation}
D = \frac{R}{3 \mathcal{M}^2}\;.
\end{equation}
where $\mathcal{M} \gg 1$ is the Mach number and $R$ is the radius of the
shell.

The zeroth order accuracy analytical solution is found by equating the rate of
change of momentum of the expanding shell to the momentum flux from the
stellar luminosity:
\begin{equation}
\label{eq:shell}
\frac{d}{dt}\left[\left(\frac{4}{3}\pi R^{3} \rho_{0}\right)\dot{R}\right] = \frac{L}{c}\;.
\end{equation}
The time evolution of the radius of the shell is therefore given by,
\begin{equation}
R = \left(\frac{3L}{2\pi c \rho_{0}}\right)^{1/4} t^{1/2}
\label{ranalytical1}
\end{equation}
We assumed here that the velocity of the particles carrying the momentum
outflow from the source, $v_\gamma$, is much larger than the shell velocity,
$\dot R$.

\noindent This solution can be further improved by taking account of the finite
speed of the wind (photon packets in our code). In this case there is a
time delay of $R/v_\gamma$ before the wind particles strike the ambient
gas. In addition, external pressure from the ambient medium provides some
restoring force, slowing down the expansion of the bubble. Factoring
these effects into equation \ref{eq:shell} leads to a more accurate
description of the shell:
\begin{equation}
\label{eq:shell_numerical}
\left[\left(\frac{4}{3}\pi R^{3} \rho_{0}\right)\dot{R}\right] = \frac{L}{c}
\left (t - \frac{R}{v_{\gamma}}\right ) - 4\pi R^{2} \rho c_{s}^2 t
\end{equation}
which is integrated numerically to obtain a solution for $R$.

\begin{figure}
\centerline{\psfig{file=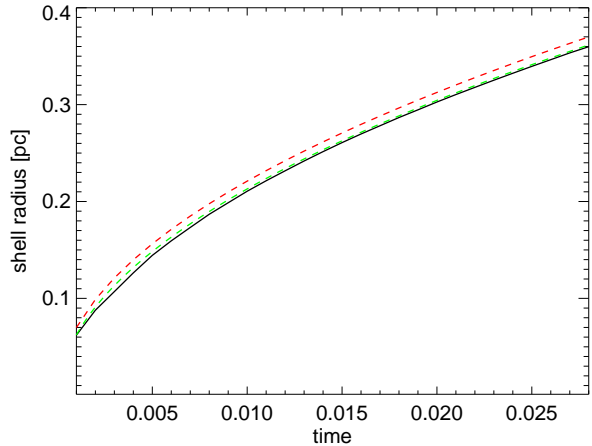,width=0.49\textwidth,angle=0}}
\caption{Radius of a shell expanding into a uniform medium as a function of
  time (see \S \ref{sec:wind}. The shell is driven by an isotropic outflow
  from a point source. The solid black line is the evolution of the density
  peak (corresponding to the position of the shock front), the red dotted line
  is the analytical solution without corrections as per equation
  \ref{eq:shell} and the dotted green line is the more accurate analytical
  solution corrected for a finite photon speed $< c$ and the pressure of the
  external medium.}
\label{fig:expanding_shell}
\end{figure}

Figure \ref{fig:expanding_shell} shows the results of the test with $p_\gamma=
10^{-7}$. The solid black line shows the time evolution of the shock front
(identified by the peak in the density distribution) in the SPH simulation.
The analytical result given by equation \ref{ranalytical1} is shown with the
red dotted line.  The green dotted line in Figure \ref{fig:expanding_shell} is
a numerical solution of the more accurate equation
\ref{eq:shell_numerical}. The difference between the two expected solutions
(red and green) is rather minimal since the external pressure is small
compared with the ram pressure of the outflow and the photon packet's velocity
is large, $v_\gamma = 200$.

Figure \ref{fig:expanding_shell} shows an excellent agreement between the
expected 1D solution and the simulation's result. 

\section{Plane-parallel slab tests}\label{sec:slab}

A simple but useful test of gas dynamics is provided by optically thin gas
illuminated by a constant radiation flux. Clearly the radiation force in this
case must be a constant independent of position or time. In the stochastic
approach of our method, however, the force can in principle vary spuriously
due to statistical fluctuations, which may lead to numerical artefacts. Our
goal here is to investigate their nature, magnitude and scaling with the
number of photon packets used.

As in \S \ref{sec:onspot}, the initial condition is a uniform density box with
a side equal to 1. The SPH particles are allowed to interact with each other
hydrodynamically only (i.e., self-gravity is turned off). In the tests
reported here, a given number of photons $N_\gamma$ propagates in the positive
$x$-direction. The momentum of photons is set by requiring the time and volume
averaged radiative acceleration be equal to a parameter $a_0$.

In the limit of an infinite number of photons and an absence of any numerical
artefacts, all the SPH particles should be accelerated with acceleration
$a_0$, so that their $x$-component of velocity is $v_x = a_0 t$. In reality
different SPH particles interact with different number of photons due to
stochastic fluctuations, and hence their accelerations and $v_x$ are not
exactly equal. We define the Mach number of spurious SPH velocity fluctuations
as a measure of the error in this test:
\begin{equation}
M_{\rm err}(t) =
\frac{1}{c_s} \left[\frac{1}{N}\;\sum_{i=0}^{i=N} ({\bf v_i-v(t)})^2\right]^{1/2}\;,
\label{macherr}
\end{equation}
where $N$ is the total number of the SPH particles, and $v(t) = v_x(t) = a_0 t$ is the
average particle velocity at time $t$. The quantity in the square brackets is
the velocity dispersion, of course. 

\begin{figure}
\centerline{\psfig{file=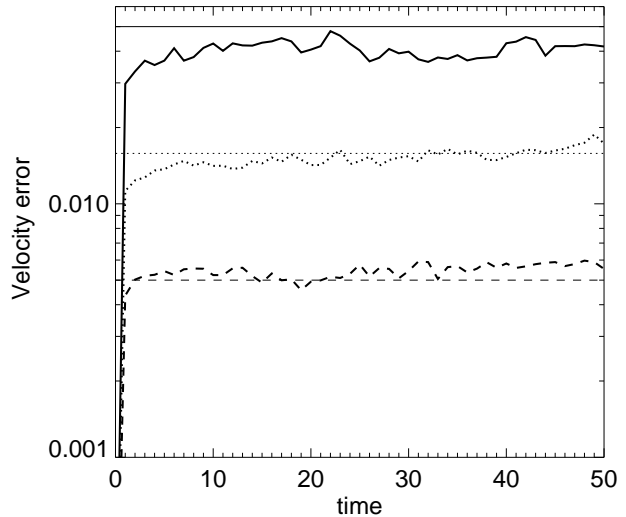,width=0.49\textwidth,angle=0}}
\caption{Velocity error defined as the Mach number of velocity fluctuations
  (equation \ref{macherr}) versus time for three optically thin slab tests
  described in \S \ref{sec:slab}.  The number of photon packets is varied from
  $30$ (upper thick solid curve), through 300 (middle, dotted curve) to $N_{\gamma} =
  3\times 10^3$ (lower, dashed curve). The respective straight lines show the
  ``theoretical'' estimate of the error given by equation \ref{dispv}.}
\label{fig:slab_box_error}
\end{figure}

We can build a simple ``theory'' to estimate the error in these tests. We
shall assume that velocity dispersion is driven by the stochastic fluctuations
in the number of photon-SPH interactions in different regions of the box. We
shall also assume that fluctuations in these will be uncorrelated on a
timescale of order the sound crossing time of the box, $L_{\rm box}/c_s$.
During this time the mean number of the interactions between photons and a
parcel of gas with size $h^3$ (where $h$ is the smoothing length) is
\begin{equation}
N_{\rm pass} \sim h^2 n_\gamma v_\gamma L_{\rm box}/c_s\;= N_\gamma
\frac{h^2}{L_{\rm box}^2} \frac{v_\gamma}{c_s}\;,
\label{npass}
\end{equation}
where $n_\gamma = N_\gamma/L_{\rm box}^3$ is the volume average density of
photons in the box.  The parcel is going to be accelerated to velocity $a_0
L_{\rm box}/c_s$ during this time. Each photon then accelerates the parcel by
a velocity increment of
\begin{equation}
\Delta v_1 = a_0 L_{\rm box}/c_s/N_{\rm pass}\;.
\label{dv1}
\end{equation}
As the number of photon-parcel interactions fluctuates by $\Delta N_{\rm
  pass} = N_{\rm pass}^{1/2}$, we expect that random velocity fluctuations
will be of the order of
\begin{equation}
\Delta v = \Delta v_1 N_{\rm pass}^{1/2}\;=  \; \frac{a_0 L_{\rm
    box}}{c_s}\;\frac{1}{N_{\rm pass}^{1/2}}\;.
\label{dispv}
\end{equation}
Figure \ref{fig:slab_box_error} compares the measured error and the one
predicted by equation \ref{dispv} for three runs. In the runs the SPH particle
number is fixed at $N_{\rm sph} = 10^4$, $a_0 = 0.1$, $v_\gamma = 10$, and the
number of photon packets inside the box is varied from $30$ (upper curve),
through 300 (middle, dashed curve) to $N_{\gamma} = 3\times 10^3$ (lower,
dashed curve). The respective straight lines show the ``theoretical'' estimate
given by equation \ref{dispv}. The agreement is reassuringly good. One can
also note that the error indeed saturates rather than grows with time, except
for a few initial sound crossing times. This is due to the stochastic
fluctuation becoming uncorrelated on dynamical time, as we assumed above.

We also ran a large number of additional tests all with the same setup
described in this section, but varying the number of SPH particles up to
$2\times 10^5$, mean radiation pressure acceleration $a_0$ (increasing it to
$1$ and $10$), and also varying the photon velocity $v_\gamma$ by a factor of
2 in either direction. The resulting errors and their scalings were compared
with that predicted by equation \ref{dispv}, confirming the excellent
agreement of the expected and measured errors further.

Equation \ref{dispv} shows that, at the same $a_0$ (equivalently, the same radiation
flux), the velocity error scales as
\begin{equation}
\Delta v \propto  \;\frac{1}{h N_\gamma^{1/2}}  \; \propto   \; \frac{N_{\rm sph}^{1/3}}{
N_\gamma^{1/2}}\;.
\label{scaling}
\end{equation}
The scaling of $\Delta v$ with the number of photon packets as $\propto
N_\gamma^{-1/2}$ could be intuitively expected on the basis of Poisson
statistics of random fluctuations. The scaling of equation \ref{dispv} with
the smoothing length $h$, which is proportional to $N_{\rm sph}^{-1/3}$, shows
another important point. If a larger number of SPH particles is used,
resulting in a higher spatial resolution within the simulation volume, then
the number of photon packets should also be increased (if one wishes to keep
the velocity errors within a given limit).

\section{A suspended optically thin cloud}\label{sec:static}

Consider optically thin gas in the vicinity of a very massive point mass,
e.g., a black hole, radiating exactly at the Eddington limit. Obviously, gas
particles should experience no net force from the point source. In particular,
if SPH particle velocities are zero initially, they should remain zero, and
any gas motion is an evidence of problems in the numerical methods used.

For ease of analysis, we consider a self-gravitating gas cloud, with a
polytrophic index $\Gamma = 5/3$. In the absence of external forces the cloud
settles into a well known equilibrium configuration for polytropic stars. This
equilibrium state is obtained by relaxing (that is evolving) the isolated
cloud for many dynamical times to provide a noise-free initial condition.

Gravity is switched on in this test. However, as our focus is on testing the
radiation pressure force, the gravitational force of the gas acting on the
point mass is switched off. The radiation source is hence fixed in its initial
location, which simplifies a quantitative analysis of the results below.

The mass unit for the test is the mass of the point source, $M$. The mass of
the cloud is $0.023$. The polytropic constant is such that the equilibrium
size of the cloud, $R_{\rm cl}$, is approximately 4 units of length. 25,000 SPH
particles are used to model the cloud.

\del{We placed the cloud near the point mass with mass $M=1$ emitting at its
  Eddington rate as described above.  The mass of the cloud is $0.023$ in
  these units, and the distance between the cloud's centre and the black hole,
  $z_0$, is larger than the initial size of the cloud, $R_{\rm cl} \approx 4$.
  Physical units employed here, although irrelevant for the single SPH
  particle case, are as follows. The unit mass is $M_U = 3.5\times 10^6
  \msun$, and the unit of length is $L=1.2\times10^{17}$~cm (these units are
  convenient for modelling the Milky Way's central parsec).}

The tidal force from the central point is of the order of $F_{\rm t} \sim 2
(GMM_{\rm cl}/R^3) R_{\rm cl}$, where $R$ is distance to the point mass. There
are two distinct regimes, then.  In the first the tidal force acting on the
cloud exceeds the self-gravity of the cloud, $F_{\rm sg} \sim (GM_{\rm
  cl}^2/R_{\rm cl}^2)$, whereas in the opposite regime the self-gravity of the
cloud dominates over the tidal force. The former regime occurs when $R=z_0$,
the separation between the point mass and the centre of the cloud, is
relatively small, $z_0\simgt R_{\rm cl}$. The case of a negligible tidal
force, $F_{\rm t}\ll F_{\rm sg}$, occurs when the cloud is far away from the
black hole, e.g., $z_0\gg R_{\rm cl}$.

\subsection{$z=10$ tests}

In these tests the cloud is initially positioned at $z=0$ and the emitter is
at $z=-10$. At this separation, the tidal force is approximately 10 times
greater than self-gravity of the cloud. Thus, a $\sim 10$\% error in the
radiation pressure force is larger than the self-gravity holding the cloud
together. These tests are hence expected to be very sensitive to numerical
deficiencies of our method.

We performed two tests in this setup. In the first, labelled LRZ10 (low
resolution, $z=10$), the dimensionless photon momentum is $p_\gamma = 2\times
10^{-8}$, whereas in the second, named HRZ10 (higher resolution), $p_\gamma =
2\times 10^{-9}$. The mass of the SPH particles is the same in both tests,
$m_{\rm SPH} \approx 10^{-6}$. The typical momentum of SPH particles, defined
as $p_{\rm SPH} = m_{\rm SPH} \sqrt{G M/z_0}$, is $ p_{\rm SPH} \approx 4
\times 10^{-7}$ in code units. Therefore, in both of these runs $p_\gamma \ll
p_{\rm SPH}$.

\begin{figure*}
\begin{minipage}[b]{.48\textwidth}
\centerline{\psfig{file=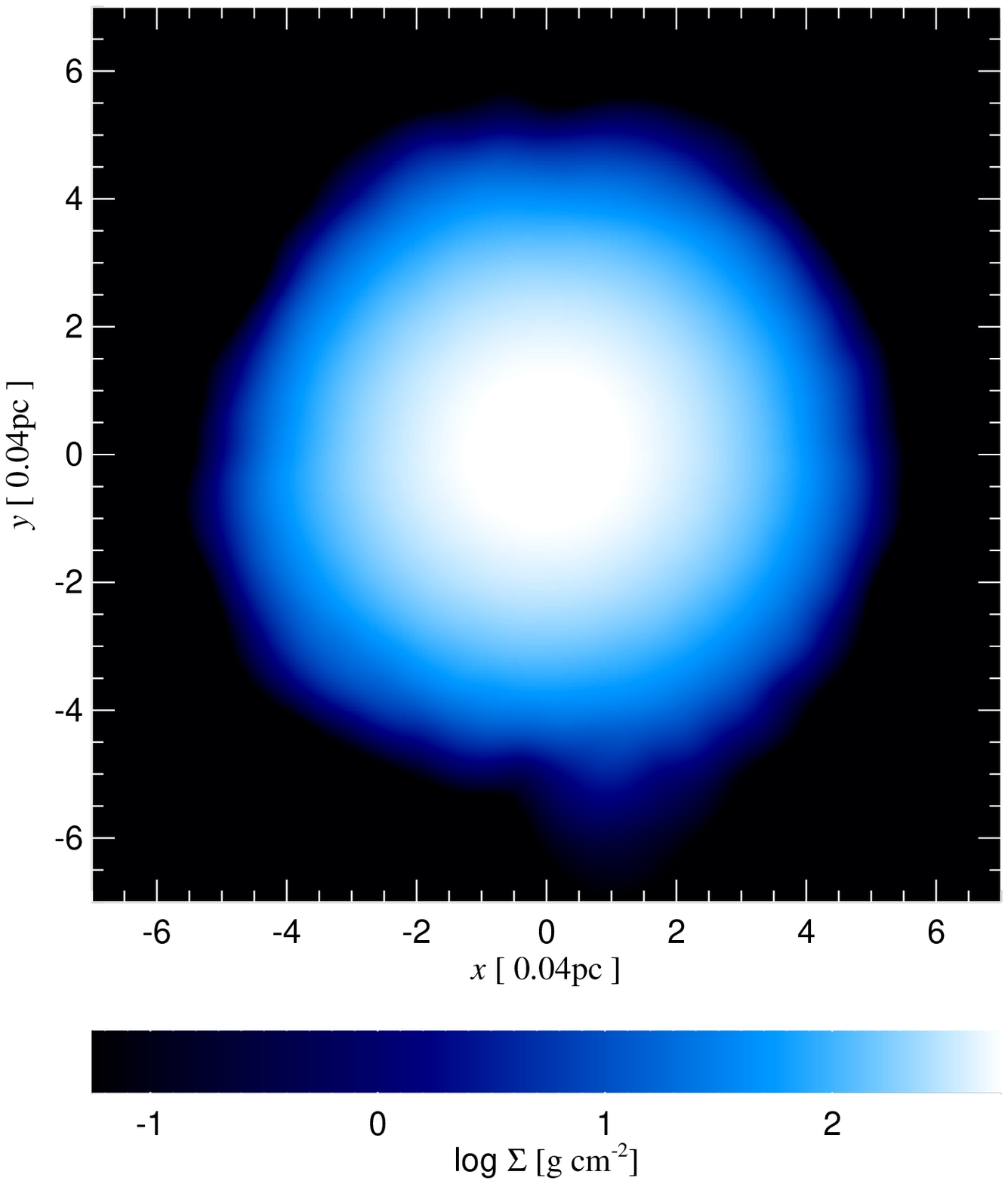,width=0.99\textwidth,angle=0}}
\end{minipage}
\begin{minipage}[b]{.48\textwidth}
\centerline{\psfig{file=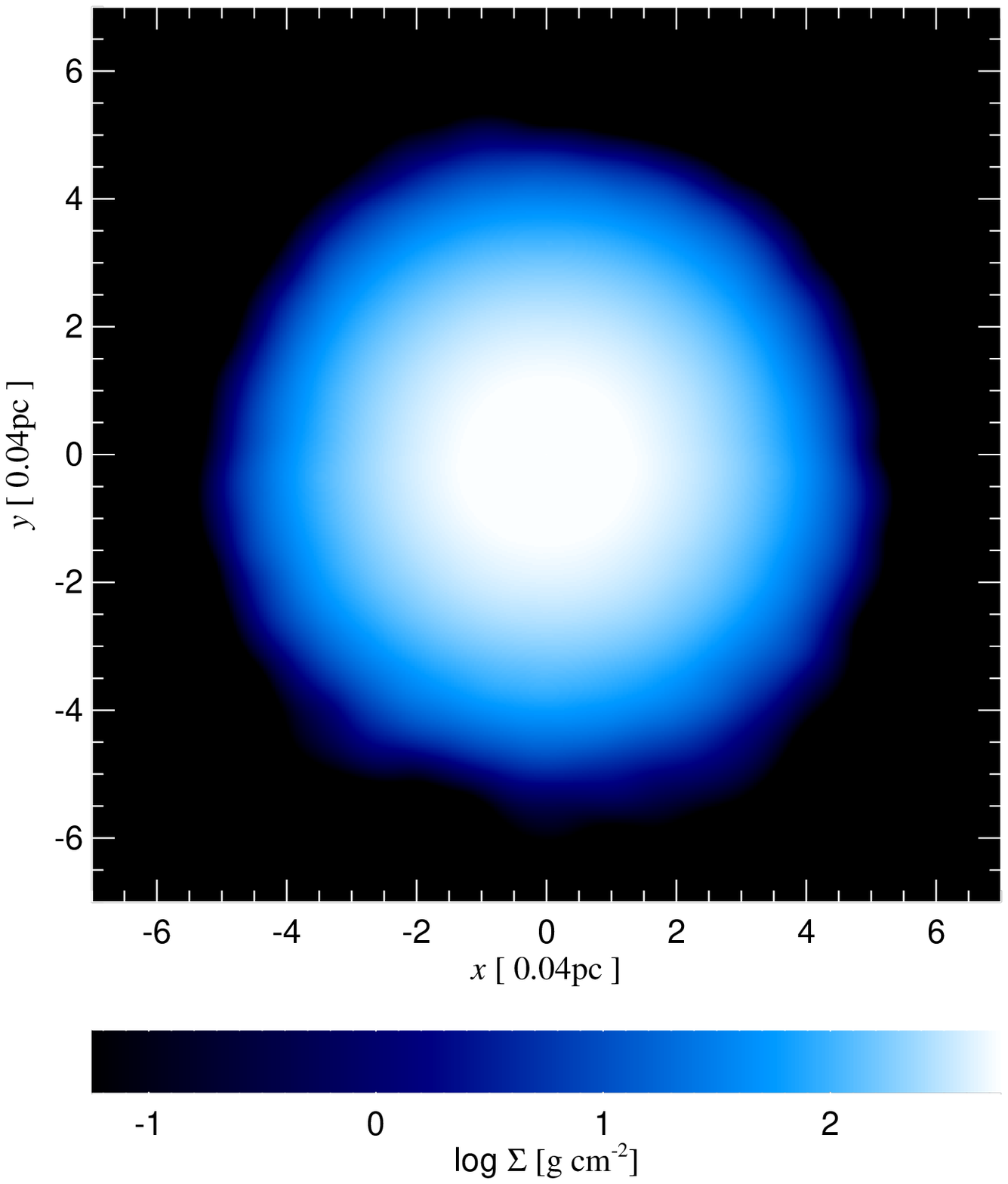,width=0.99\textwidth,angle=0}}
\end{minipage}
\caption{Column density of gas in the tests LRZ10 (left panel) and HRZ10 (right
  panel).  The optically thin cloud in the tests experiences gravity from the
  central source at $z=-10$ and the radiation pressure force at the Eddington
  limit. The two forces should exactly balance each other. The snapshots are
  for dimensionless time $t=100$, roughly three free-fall times. The higher
  resolution test HRZ10 (lower $p_\gamma$) shows a smaller degree of cloud
  deformation, as expected.
}
\label{fig:z10_standard}
\end{figure*}

Figure \ref{fig:z10_standard} shows the column density profile of the cloud at
time $t=100$ for both runs. The left panel shows run LRZ10 and the right panel
shows HRZ10. Time $t=100$ corresponds to about 3 free-fall times at the given
source-cloud separation. In both cases there is a certain deformation of the
cloud. As expected, the lower resolution test produces poorer results than the
higher resolution test. A more detailed error analysis is described in \S
\ref{sec:error_analysis} below.

\subsection{z=30 test}

In this test the separation of the cloud and the source is set to $z_0 = 30$,
thus we label the test Z30.  The photon packet momentum is $p_\gamma =
10^{-8}$ in the test. The parameters of the gas cloud are exactly the same as
in tests LRZ10 and HRZ10, but because the cloud is farther away from the point
mass, typical SPH momentum is a little lower, $p_{\rm SPH} \sim 2.5 \times
10^{-7}$. Thus in terms of ratio $p_\gamma/p_{\rm SPH}$ test Z30 is very
similar to LRZ10.

However, since the cloud is farther away, the self-gravity of the cloud
actually exceeds the tidal force by a factor of about 3. Therefore, one
expects intuitively that the same level of stochastic error in the radiation
pressure force will result in smaller perturbations as the radiation pressure
force itself is smaller. This is indeed borne out by the test. By the time
$t=500$, which is again about three free-fall time scales from the clouds'
initial location, no visible deformation of the cloud has occurred. We
therefore do not show the column density for test Z30.

\subsection{Error analysis}\label{sec:error_analysis}

\begin{figure}
\centerline{\psfig{file=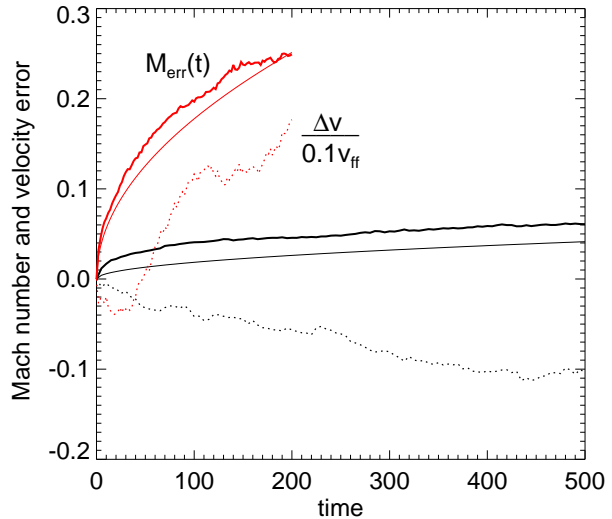,width=0.49\textwidth,angle=0}}
\caption{The Mach number of numerical fluctuations (thick solid curves) and
  the average SPH velocity in units of 0.1 free-fall velocity (dotted) for
  tests HRZ10 (red curves) and test Z30 (black curves). Theoretically
  estimated Mach number errors are also shown with thin solid curves of same
  colours. }
\label{fig:z10_standard_err}
\end{figure}

To quantify the accuracy of the tests, we calculate the centre of mass
velocity of the cloud, $v_{\rm cm}$ as a function of time. Using this in place
of ${\bf v(t)}$ in equation \ref{macherr}, we define the mean Mach number of
spurious velocity fluctuations in the cloud.  Figure
\ref{fig:z10_standard_err} shows these quantities as a function of time for
LRZ10 (red curves) and Z30 (black) runs. The Mach number of velocity
fluctuations are shown with thick solid lines. The centre of mass
velocities are shown normalised to $0.1 v_{\rm K}$, where $v_{\rm K} =
\sqrt{GM/z_0}$, the circular Keplerian velocity at the appropriate separation
$z_0$ for the test.

The motion of the cloud's centre of mass characterises the systematic error
of the method.  Both runs demonstrate that the velocity of the centre of mass
of the cloud is between 1 to 2 \% of the Keplerian velocity after about 3
free-fall times. This translates into inaccuracy in the calculation of the
radiation pressure force versus point mass gravity, averaged over all SPH
particles, of a fraction of a percent. This accuracy should be sufficient for
simulations in which gravity forces are calculated to a similar fractional
precision. Furthermore, we found that this offset, although always small for a
large enough number of photon packets, actually depends on the random number
generator used to approximate the isotropic photon field. The results can thus
be improved by using quasi-random rather than pseudo-random sequences for
photon packet's angular distribution. We shall investigate this issue in the
future.

The Mach number of the fluctuations, instead, quantifies the difference in the
radiative accelerations received by different parts of the cloud. We again use
the simple logic of random stochastic fluctuations in the number of photon
packets $N_{\rm pass}$ interacting with an SPH particle with smoothing length
$h$, as explained in \S \ref{sec:slab}. Thin solid curves in Figure
\ref{fig:z10_standard_err} show the Mach number of velocity errors predicted
by this simple argument. Evidently, the errors are again explainable by the
Poisson noise estimates.

\section{A spherically symmetric accretion test }\label{sec:accretion}

In this test the initial condition is the same optically thin cloud as used in
\S \ref{sec:static}, but the point mass source is located exactly at the
centre of the cloud, and allowed to accrete gas particles that are separated
from it by less than $R < R_{\rm acc} = 0.2$. The radiation field in this test
is halved to $L=0.5\; L_{\rm Edd}$. While the analytical solution to this
problem (for the given initial density profile) is not known to us, we note
that the setup is exactly identical to a non-radiating sink particle with mass
$M=1/2$ placed into the centre of the cloud and allowed to accrete the gas
within the sink radius $R_{\rm sink}$. The desired control result is thus
obtained by running the hydro-gravity part of the code only with the sink
particle mass $M=1/2$. Photon momentum is set to $p_\gamma = 10^{-8}$.

\begin{figure}
\centerline{\psfig{file=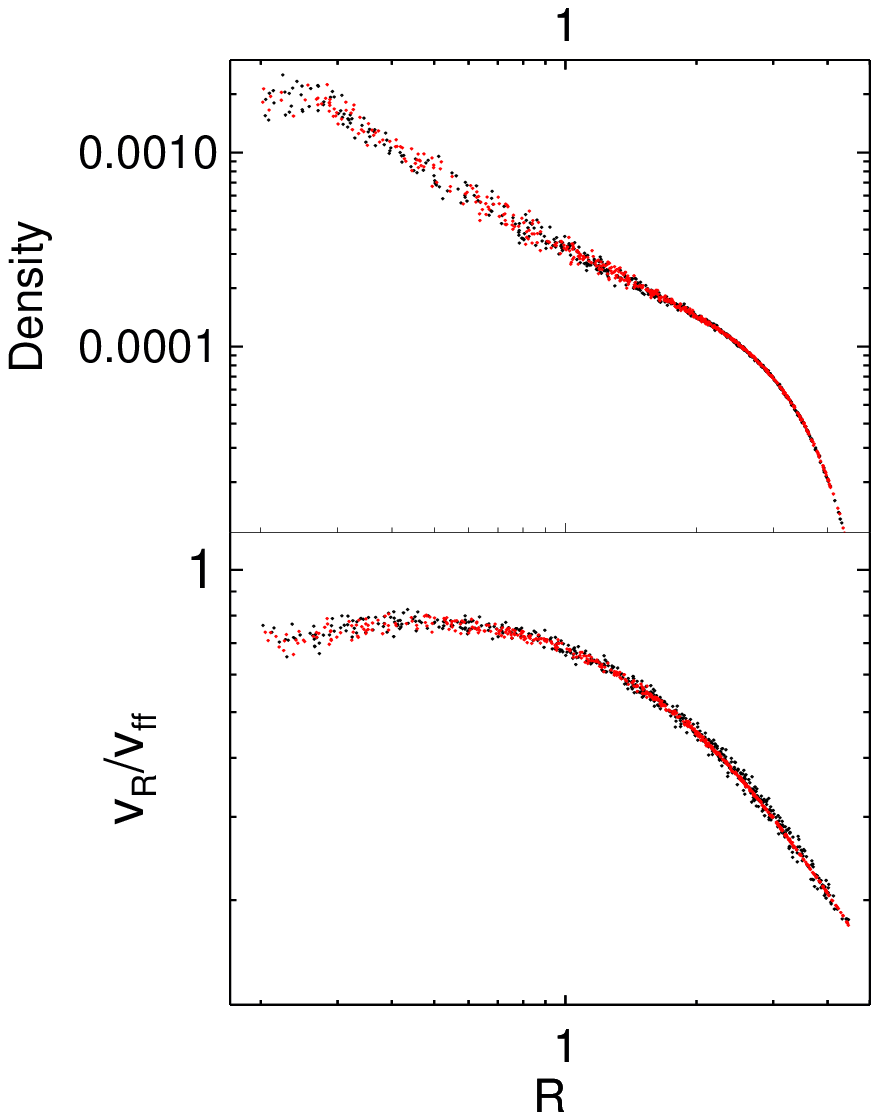,width=0.49\textwidth,angle=0}}
\caption{Spherically symmetric accretion test on the central source of mass
  $M=1$ emitting at exactly $1/2$ of the Eddington limit. The upper and the
  lower panels show the profiles of the SPH particle density and the ratio of
  radial velocity to the local free-fall velocity, respectively. The black
  dots show the results of the radiation transfer $L=1/2 L_{\rm Edd}$
  simulations, whereas the red ones show the results of accretion onto a
  non-radiating point mass with mass equal to $1/2$. The two tests should
  yield identical results.}
\label{fig:accretion1}
\end{figure}

Figure \ref{fig:accretion1} presents both the radiation hydrodynamical
simulation and the control hydro run with $M=1/2$ at dimensionless time
$t=3.5$. The lower panel shows the SPH particle radial velocities normalised
by the local free-fall velocity $v_{\rm ff} = \sqrt{GM/R}$ with $M=1/2$. The
upper panel shows the density profile of the cloud. Both velocity and density
profiles of the RHD simulation and the control run are nearly identical,
although one can notice a larger amount of scatter in the $L= 0.5\; L_{\rm
  Edd}$ test. The results continue to be very similar at latter times as well,
and the accretion rate histories are also very similar. The radiation transfer
method thus performs reasonably well.

To quantify the errors of the tests, we estimate their magnitude based on the
Poisson noise arguments in the same manner as in \S \ref{sec:slab}. In
particular, we can apply equations \ref{dv1} and \ref{dispv}, except that the
characteristic time is now not $L_{\rm box}/c_s$ but the local dynamical time,
$t_{\rm dyn}(R) = R^{3/2}/(GM)^{1/2}$ with $M= 1/2$ (and $G=1$ in the code
units). Thus, the number of photon packets passed through a given particle is
estimated as $N_{\rm pass}(R) = \dot{N}_\gamma t_{\rm dyn} (h/2R)^2$.  The
stochastic velocity fluctuations due to the finite number of photon packets is
then
\begin{equation}
\Delta v \sim \Delta v_1 N_{\rm pass}^{1/2}(R)\;.
\label{dv_accretion}
\end{equation}

\begin{figure*}
\centerline{\psfig{file=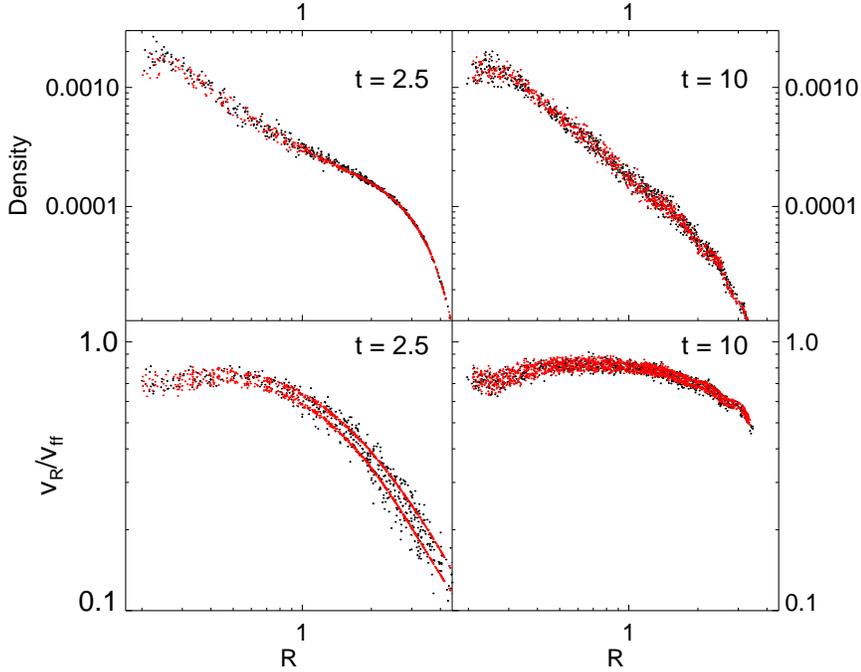,width=0.7\textwidth,angle=0}}
\caption{Similar to Figure \ref{fig:accretion1}, but for a test with $p_\gamma
  = 10^{-7}$, i.e., lower resolution in photon packets. The red dots in the
  lower panels indicate the estimated range of errors (see text in \S
  {sec:accretion}), which agrees reasonably well with the spread in the black
  dots.}
\label{fig:accretion_errors}
\end{figure*}

To compare the expected fluctuations given by equation \ref{dv_accretion} with
those actually incurred, we ran a poorer resolution test with $p_\gamma =
10^{-7}$, i.e., ten times larger than the test shown in Figure
\ref{fig:accretion1}. The errors (scatter) are then obviously larger. Figure
\ref{fig:accretion_errors} shows the results in black dots for two snapshots
at times indicated. The red lines show the control run, as before, but this
time velocity curves in the lower panel of the figure include the expected
errors. In particular, the upper red sequence of dots is $v_{\rm control} +
\Delta v$, whereas the lower red dots is $v_{\rm control} - \Delta v$, where
$\Delta v$ is calculated with equation \ref{dv_accretion}.

Evidently, the scatter in the radiation transfer simulation (black dots) is
very similar in magnitude to what is predicted by equation
\ref{dv_accretion}. At the earlier time snapshot, shown on the left side of
Figure \ref{fig:accretion_errors}, the measured velocity scatter is somewhat
larger than our prediction. We believe this is due to the fact that equation
\ref{dv_accretion} does not include the Poisson noise resulting from the
initial distribution of photons packets at $t=0$. The latter is generated by
sampling a uniform random distribution in both emission time and directions.

In conclusion, these tests demonstrate the potential of the method for
accretion and radiation pressure problems, and that equation
\ref{dv_accretion} once again provides a reliable estimate of the stochastic
errors of the results.

\section{Discussion}\label{sec:conclusion}
A new method for implementation of radiation pressure force in an SPH code has
been presented. Radiation is modelled via a time-dependent Monte-Carlo
approach. Photon packet's momentum is transferred to SPH particles via direct
photon-to-gas interactions that are calculated as photons propagate through
the field of SPH particles. The SPH density field is used to calculate the SPH
density at photon packet's positions, thus maximising the accuracy of the
method. As a result, local and global momentum conservation in the
interactions between radiation and gas is achieved.

Several tests have been presented to check the consistency and accuracy of the
method, as well as error scalings with parameters of the problem. It was found
that the main deficiency of the method, as for any Monte Carlo method, is the
Poisson noise resulting from a finite number of photon packets used. In
order to reduce velocity fluctuations to subsonic levels, photon packet's
momentum $p_\gamma$ should be chosen to be below the typical SPH particle
momentum, $m_{\rm sph} v_{\rm sph}$, where $v_{\rm sph}$ is the
characteristic SPH particle velocity.  

We shall now try to generalise and summarise the error estimates and various
constraints on performance of the code.

\subsection{Velocity errors}\label{sec:mom}

We first assume an optically thin case. Let ${\cal F}_{\rm rad}$ be the radiation
flux at a given location in the gas. The flux is related to the photon
number density $n_\gamma$, photon momentum $p_\gamma$ and photon velocity
$v_\gamma$ through
\begin{equation}
{\cal F}_{\rm rad} \sim n_\gamma v_\gamma p_\gamma c\;,
\label{fluxrad}
\end{equation}
as the energy carried by the photon packet is $p_\gamma c$. The rate at which 
photons are passing through SPH particles is
\begin{equation}
\dot N_{\rm pass} \sim n_\gamma v_\gamma \pi h^2 \sim \pi h^2 \frac{{\cal F}_{\rm rad}
}{p_\gamma c}\;.
\label{dotnpass}
\end{equation}
The momentum passed to the SPH particle by one photon packet is
\begin{equation}
m_{\rm sph} \Delta v_1 \sim \kappa \frac{m_{\rm sph}}{\pi h^2}
p_\gamma\;,
\label{dmom}
\end{equation}
and thus the velocity change $\Delta v_1 = \kappa p_\gamma/(\pi h^2)$ is
independent of SPH mass $m_{\rm sph}$ (in the optically thin
approximation). Therefore, at time $t$ the velocity fluctuations can be
estimated as
\begin{equation}
\Delta v(t) \sim \Delta v_1 \left[\dot N_{\rm  pass} t\right ]^{1/2} \sim
\frac{\kappa}{h} \left[\frac{{\cal F}_{\rm rad} p_\gamma t}{c}\right]^{1/2}\;,
\label{dvoft}
\end{equation}
 Now, let us define radiation acceleration time $t_{\rm rad} = v_{\rm
   sph}/a_{\rm rad}$, where $a_{\rm rad} = F_{\rm rad}/m_{\rm sph}$ is the
 radiative acceleration acting on the SPH particle, and $F_{\rm rad} = m_{\rm
   sph} \kappa {\cal F}_{\rm rad}/c$ is the radiation pressure force on the
 SPH particle.  During this characteristic time, the radiation pressure force
 would accelerate the SPH particle from rest to velocity $v \sim v_{\rm sph}$
 in the absence of other forces.

Using equation \ref{dvoft}, the Mach number of the fluctuations, defined by
equation \ref{macherr}, can be shown to be
\begin{equation}
M_{\rm err}(t) \sim \frac{\Delta v(t)}{v_{\rm sph}} \sim \left[\frac{\kappa
    p_\gamma}{v_{\rm sph} \pi h^2}\; \frac{t}{t_{\rm rad}}\right]^{1/2}\;.
\end{equation}
Finally, introducing the ``optical depth'' of the SPH particle as $\tau_{\rm
  sph} = \kappa m_{\rm sph}/(\pi h^2)$, we arrive at
\begin{equation}
M_{\rm err}(t) \sim \left[ \tau_{\rm sph} \frac{p_\gamma}{p_{\rm sph} } \right]^{1/2} \left(\frac{t}{t_{\rm rad}}\right)^{1/2}\;,
\label{merrfinal}
\end{equation}
where $p_{\rm sph} = m_{\rm sph} v_{\rm sph}$. This was derived in the
optically thin limit, i.e., when $\tau_{\rm sph} \ll 1$. In the limit
$\tau_{\rm sph} > 1$, the momentum passed from one photon to an SPH particle
is reduced from the expression given by equation \ref{dmom} to
\begin{equation}
m_{\rm sph} \Delta v_1 \sim p_\gamma N_{\rm nb}^{-1}\;,
\label{dmomthick}
\end{equation}
where $N_{\rm nb}$ is the number of SPH neighbours for the photon. As the
latter is always at least one when a photon packet interacts with gas, we
arrive at the following estimate for the optically thick case, 
\begin{equation}
M_{\rm err}(t) \simlt \left[\frac{p_\gamma}{p_{\rm sph} } \right]^{1/2} \left(\frac{t}{t_{\rm rad}}\right)^{1/2}\;,
\label{merrthick}
\end{equation}

Both the optically thin limit (equation \ref{merrfinal}) and the optically
thick limit (equation \ref{merrthick}) demonstrate the importance of choosing
photon packet momentum, $p_{\gamma}$ to be significantly smaller than that of
an SPH particle to guarantee a reasonable accuracy.

We also notice that applicability of the method to a particular problem
depends on the desired level of accuracy. For example, tidal disruption of a
gaseous cloud near a luminous super-massive black hole is a highly dynamic
process. To get an insight in the overall dynamics of the process, it is
sufficient to simulate the system for a few dynamical times at the peri-centre
of the orbit. A sufficiently small photon packet's momentum, $p_\gamma \sim
0.01-0.1 p_{\rm sph}$, should provide a sufficient accuracy in this
case. However if one is interested in a secular evolution of a system, then
the required precision is much higher, and careful tests should be done to
establish the required value of $p_\gamma$, or equivalently the photon
injection rate $\dot N_\gamma$.

\subsection{Number of photon packets}\label{sec:number}

Let us now estimate the required number of photon packets for a given
luminosity of the system $L_\gamma$, subject to the constraint $p_\gamma =
p_{\rm sph}/\Gamma$, $\Gamma \gg 1$. Number of packets emitted per unit time
is
\begin{equation}
\dot N_\gamma= \frac{L_\gamma}{c p_\gamma}= \frac{\Gamma
  L_\gamma}{c}\frac{N_{\rm sph}}{M_{\rm gas} v_{\rm sph}}\;,
\label{ndot_req}
\end{equation}
where we exploited the fact that $p_{\rm sph} = (M_{\rm gas}/N_{\rm sph})
v_{\rm sph}$. $M_{\rm gas}$ here is the total gas mass of the system. The
minimum time that the simulation should last for is the dynamical time,
$R/v_{\rm sph}$. During this time the number of photon packets emitted is
$N_\gamma = \dot N_\gamma R/v_{\rm sph}$, and thus
\begin{equation}
\frac{N_\gamma}{N_{\rm sph}}= \Gamma \frac{L_\gamma (R/c)}{M_{\rm sph} v_{\rm
    sph}^2} = \Gamma \frac{ E_\gamma}{E_{\rm sph}}\;,
\label{nratio}
\end{equation}
where $E_\gamma = L_\gamma R/c$ is the radiation energy emitted during the
system light crossing time and $E_{\rm sph} = M_{\rm gas} v_{\rm sph}^2$ is
the total gas energy. If there are no additional sources of radiation in the
cloud, such as bright stars, then the luminosity of the system is of the order
of $L_\gamma \sim E_{\rm sph}/t_{\rm cool}$ where $t_{\rm cool}$ is the
cooling time. In that case the ratio is
\begin{equation}
\frac{N_\gamma}{N_{\rm sph}}= \frac{\Gamma R}{c t_{\rm cool}}\;.
\label{nratio2}
\end{equation}
Many astrophysically interesting problems are in the regime where the cooling
time is comparable with dynamical time of the system, in which case the ratio
becomes 
\begin{equation}
\frac{N_\gamma}{N_{\rm sph}}= \frac{\Gamma v_{\rm sph}}{c}\;.
\label{nratio3}
\end{equation}
From these expressions it is obvious that the number of photons does not need
to exceed the number of sph particles while satisfying the $M_{\rm err} \ll 1$
condition for these particular cases. However, if there is a very bright point
source dominating the luminosity of the system, then the number of photon
packets needs to be correspondingly higher. For example, in the tests
presented in \S \ref{sec:static}, photons produced during the duration of the
simulations outnumbered SPH particles by factors of $10-100$.

\subsection{Optical depth of the system}\label{sec:odepth}

Our radiation transfer method unfortunately becomes very expensive in
optically thick medium, i.e., when the optical depth of the system, $\tau$, is
very high. This is due to two factors. First of all, photons scatter or get
absorbed/re-emitted $\sim \tau^2$ (that is many) times before they exit the
system. They thus ``hang around'' for longer before exiting the system, hence
increasing computational cost of the simulation. Secondly, they require
smaller time steps as we now show.

Recall that in \S \ref{sec:prompt} we argued that it is numerically
permissible and beneficial to use photon packets with velocity $v_\gamma$
smaller than the velocity of light in the following case. In moderately
optically thick, non relativistic plasmas, photons pass through the system
much quicker than dynamical time, and hence exact value of $v_\gamma$ is
unimportant. The photon time step can then be increased inversely proportional
to $v_\gamma$.  However, this is possible only as long as
\begin{equation}
v_\gamma  \gg v_{\rm sph} \left(1 + \tau\right)\;,
\label{vgamma}
\end{equation}
or else using too small a value for $v_\gamma$ would incorrectly put the
radiation transfer into the dynamic diffusion limit (see \S
\ref{sec:prompt}). Further, in the optically thick case the photon time step is
limited to $\Delta t_\gamma = \delta_t \lambda/v_\gamma$, where $\lambda$ is
the mean free path (equation \ref{dtgamma}). Comparing this to dynamical time
of the system,
\begin{equation}
\frac{\Delta t_\gamma}{t_{\rm dyn}} = \frac{\delta_t \lambda}{R} \frac{v_{\rm
    sph}}{v_\gamma} \ll \frac{\delta_t}{\tau (1 + \tau)}\;,
\label{dtcomp}
\end{equation}
where we used equation \ref{vgamma} to constrain $v_\gamma$. This expression
changes to
\begin{equation}
\frac{\Delta t_\gamma}{t_{\rm dyn}} =  \frac{\delta_t v_{\rm sph}}{\tau c}
\label{dtcomp2}
\end{equation}
if one sets $v_\gamma = c$. Obviously, in both of these cases the time step
becomes very small as $\tau$ increases, and the code becomes quite
inefficient.

In problems with a very high optical depth combining the dynamic Monte Carlo
and the diffusion approximation in very optically thick regions would be
optimal, although it is not clear whether such ``on the fly'' method could be
devised.

\section{Conclusion}

A new time-dependent algorithm to model radiation transfer in SPH simulations
was presented. In the present paper we concentrated on the radiation pressure
effects, i.e., the momentum transfer between the radiation and the gas,
assuming a given equation of state for the gas. We performed a number of tests
of the code. 

Our method is grid-less and can be applied to arbitrary geometries.  The main
disadvantage of the method, as with any photon packet based schemes, is the
Poisson noise due to a finite number of photons. On the other hand, these
stochastic errors can be readily estimated and controlled by increasing the
number of photon packets. The method is best applicable to optically thin or
moderately optically thick systems, as following photon trajectories become
very expensive at high optical depths.

Extension of the approach to include energy exchange between the gas and
radiation will be presented in a future paper. Multi-frequency radiation
transfer can also be naturally added. 

\section{Acknowledgments}

The authors express gratitude to Volker Springel who provided the code
Gadget-3, helped with the numerical implementation and provided useful
comments on the paper.  Theoretical astrophysics research at the University of
Leicester is supported by a STFC Rolling grant.

\bibliographystyle{mnras}

\label{lastpage}

\end{document}